\documentclass[amsmath,amsfonts,twocolumn,showpacs,prl]{revtex4}
\usepackage{graphicx}
\usepackage{epsfig}
\usepackage{bbm}
\usepackage{bm}
\usepackage{dcolumn}
\usepackage{amsmath}
\usepackage{amssymb}
\usepackage{color}
\usepackage[varg]{txfonts}

\newcommand{\beg}{\begin{equation}}
\newcommand{\en}{\end{equation}}
\newcommand{\bp}{\mathbf p}
\newcommand{\bq}{\mathbf q}
\newcommand{\bk}{\mathbf k}
\newcommand{\br}{\mathbf r}

\newcommand{\bn}{\mathbf n}

\newcommand \bel  {\begin{align}}
\newcommand \enl  {\end{align}}

\newcommand{\veps}{\varepsilon}
\newcommand{\eps}{\epsilon}

\newcommand{\dg}{^\dagger}

\begin{document}

\title{Inverse Faraday effect in disordered two-dimensional electronic systems}

\author{Maxim Dzero}
\affiliation{Department of Physics, Kent State University, Kent, Ohio 44242, USA}


\date{\today}

\begin{abstract}
I formulate a theory of the inverse Faraday effect in impure two-dimensional metallic system with lifted spin degeneracy induced by Rashba spin-orbit coupling. Using the formalism of non-equilibrium quantum field theory, the static contributions to the current density up to the second order in powers of external electromagnetic field are evaluated. For circularly polarized light one of the contributions describes the emergence of static magnetization. It is  shown that the direction of induced magnetization may change depending on the frequency of the external radiation and disorder scattering rate. I also find that at large frequencies the leading contribution to the  induced magnetization is proportional to the square of spin-orbit interaction and is inversely proportional to the fifth power of the frequency of an external electromagnetic wave.
\end{abstract}

\maketitle

\paragraph{Introduction.}
Almost sixty years ago Van der Ziel, Pershan and Malmstrom\cite{IFE-Exp-1965} have experimentally observed optically induced magnetization in an non-absorbing ceramic material Eu$^{2+}$:CaF$_2$. By elucidating the connection between this intriguing discovery and the Faraday effect, which consists in the rotation of the light polarization in constant magnetic field, the authors of Ref. \cite{IFE-Exp-1965} introduced the term 'inverse Faraday effect' (IFE) which meant to describe the appearance of magnetization under the influence of the circularly polarized light. It is worth mentioning that the microscopic physical mechanism governing the inverse Faraday effect has been theoretically discovered by Pitaevskii \cite{Pit1961} who showed that an application of a variable circularly polarized light to an optically transparent and dispersive medium produces an average magnetic moment. IFE is a nonlinear effect for the magnitude of the induced static magnetic moment must be proportional to the intensity of the incoming wave \cite{Pit1961}.

Since then the IFE has been studied extensively in the variety of physical systems  \cite{Pershan1966,Edelstein1998,Battiato2014,yang2022inverse,mou2023reversed,mou2023chiral,Putilov2023-IFESC,Mironov2021-IFESC,Buzdin2023,Parafilo2022Fl,Croitoru2023,IFE-Mott2022,gao2020topological,parchenko2023plasmonenhanced,Han_2023,Balatsky2023,dzero2024ife}. This is hardly surprising for, as it has been recently emphasized by several authors \cite{Buzdin2023,Croitoru2023,Balatsky2023}, IFE possesses necessary properties which can be utilized for the ultrafast manipulation of magnetic states in various modern device applications. Apart from the engineering aspects this effect also presents an interest from the fundamental physics point of view.  For example, it has been recently shown that IFE in conventional superconductors may also encode the signatures of the collective modes associated with the fluctuations of the superconducting order parameter $\Delta$ and, as a consequence, the frequency dependence of the induced magnetization is predicted to develop a minimum at $\omega_{\textrm{min}}=2\Delta$ corresponding to the resonant excitation of an amplitude (Schmid-Higgs) mode \cite{Kulik1981,dzero2024ife}.

Among several microscopic mechanisms for realization of IFE in metals, perhaps the most intuitive one involves the spin-orbit coupling interaction \cite{Edelstein1998,TaguchiTatara2011,Juraschek2017,AxialMagneto2021}.  Indeed, the spin-orbit coupling provides the mechanism for transferring the angular momentum of the circularly polarized light to the orbital motion of the electrons and then to their spin. For example, in two-dimensional electronic system without inversion center Rashba spin-orbit coupling enters into the Hamiltonian as a momentum dependent magnetic field. With accuracy up to the linear order in external electric field it gives rise to spin polarization dependent correction to the distribution function $\delta n_\eps\propto \alpha_{\textrm{so}}\tau_s(({{\mathbf c}\times{\mathbf E}})\cdot{\mbox{\boldmath $\sigma$}})\partial_\eps n_\eps$ where $\alpha_{\textrm{so}}$ is the strength of the spin-orbit coupling, $\tau_s$ is the Dyakonov-Perel spin-relaxation time, $n_\eps$ single particle distribution function, ${\mathbf c}$ is the unit vector perpendicular to the plane of motion and ${\mathbf E}$ is an external electric field \cite{MSH2004,Andrey2006}. Consequently, one may argue that by going to the next order in powers of electric field one then should find the corresponding correction to the distribution function which, after being substituted into expression for the current density, ultimately accounts for the emergence of the static spin polarization \cite{Edelstein1998}.

Earlier works on the IFE in metallic systems have incorporated the spin-orbit coupling explicitly to evaluate the dependence of the induced magnetization on frequency of external light and strength of the spin-orbit interaction. For example, the magnitude of the IFE has been found by evaluating Feynman diagrams \cite{Edelstein1998,TaguchiTatara2011}. Specifically, Edelstein \cite{Edelstein1998} has considered disordered two-dimensional Fermi gas with broken mirror symmetry which was accounted for by Rashba spin-orbit coupling and found that for the high frequency light $\omega\tau\gg 1$ and weak spin orbit coupling $\zeta=\alpha_{\textrm{so}}k_F\tau\ll 1$ static magnetization is proportional to $(\omega\tau)^{-3}$, where $\tau^{-1}$ is the relaxation rate due to disorder scattering and $k_F$ is the Fermi momentum. What is particularly surprising about this result is that it vanishes in the limit $\tau\to\infty$ indicating the absence of the IFE in a clean system. I also note that since $\tau_s^{-1}\approx \alpha_{\textrm{so}}k_F\zeta/(1+4\zeta^2)$ the limit $\zeta\ll 1$ corresponds to very long Dyakonov-Perel relaxation time, $\tau_s\gg\tau$, which according to the results of Ref. \onlinecite{Edelstein1998} implies that the induced magnetization becomes independent of the strength of the spin-orbit coupling.  Almost a decade later this problem was revisited  by Taguchi and Tatara who treated the spin-orbit interaction by perturbation theory and found the induced magnetization to be linearly proportional to both spin orbit coupling and frequency \cite{TaguchiTatara2011}. Thus, it appears that the results of these two papers are clearly at odds with each other despite the fact that the same model Hamiltonian and methodology have been used to obtain the frequency dependence of the induced magnetization. Such state of affairs is hardly satisfactory and therefore needs to be resolved. 

In this Letter I formulate a microscopic theory of inverse Faraday effect for the two-dimensional electronic system with broken inversion symmetry. Given the essentially non-equilibrium nature of the effect, one necessarily needs to adopt the Keldysh quantum field theoretical approach \cite{Kamenev2011}. In what follows below the quantum kinetic equation for the Wigner distribution function is derived and then solved by perturbation theory in powers of the electric field. The results, which have been obtained for the model with linear-in-momentum spin-orbit coupling and local disorder potential, are applicable for the arbitrary strength of the spin-orbit interaction and strength of disorder potential provided $\veps_F\tau\gg 1$ is satisfied ($\veps_F$ is the Fermi energy). At high frequencies $\omega\tau\gg 1$ I find that the induced magnetization is proportional to the square of the spin-orbit coupling and inversely proportional to the fifth power of frequency. In the case of small frequencies $\omega\tau\ll 1$, the static magnetization is found to be linearly proportional to frequency. These results are also valid in the fairly broad range of values of parameter $\zeta$. Most importantly,  I find  that the direction of the induced magnetization ${\mathbf M}_{\textrm{ind}}$ can be switched depending on the value $\omega$ (and dimensionless parameter $\zeta$), i.e. ${\mathbf M}_{\textrm{ind}}\to -{\mathbf M}_{\textrm{ind}}$ as a function of $\omega$. Such effect can be interpreted as being governed by the difference in the chirality of the underlying spin-orbit split bands which determine the magnitude and direction of the induced nonlinear current. 

\paragraph{Model and formalism.}
The model Hamiltonian describes the non-interacting electrons which are constrained to move in a plane and are subjected to external time dependent vector potential, Rashba spin-orbit coupling and disorder potential:
\beg\label{Eq1}
\begin{split}
\hat{\cal H}=\frac{1}{2m}\left[\hat{\bp}-\frac{e}{c}{\mathbf A}(\br,t)\right]^2+\alpha_{\textrm{so}}\hat{\mbox{\boldmath $\eta$}}\cdot\left(\hat{\bp}-\frac{e}{c}{\mathbf A}(\br,t)\right)+U(\br).
\end{split}
\en
Here $\hat{\mbox{\boldmath $\eta$}}=\left({\vec e}_z\times\hat{\mbox{\boldmath $\sigma$}}\right)$, $\hat{\bp}=-i{\mbox{\boldmath $\nabla$}}$ is electron momentum operator, $m$ is electron mass, ${\vec e}_z$ unit vector along the $z$-axis which coincides with the normal to the plane of motion, $U(\br)$ is a disorder potential and ${\mathbf A}(\br,t)$ is a vector potential of the periodically modulated electric field ${\mathbf E}(\br,t)=-(1/c)\partial_t{\mathbf A}(\br,t)={\mathbf E}_0e^{i(\bq\br-\omega t)}+{\mathbf E}_0^*e^{-i(\bq\br-\omega t)}$.
In what follows I focus on the case of a circularly polarized electromagnetic wave, 
${\mathbf E}_0=E_0({\vec e}_x+i{\vec e}_y)$ (here ${\vec e}_{x,y}$ are the unit vectors along the $x$ and $y$ axes correspondingly). As it will be shown below only in this case does one find the nonzero static magnetization $\propto |{\mathbf E}_0|^2$ for the model defined by the Hamiltonian (\ref{Eq1}). The disorder is assumed to be local and is described by a correlation function
$\left\langle U(\br)U(\br')\right\rangle_{\textrm{dis}}={\delta(\br-\br')}/({2\pi\nu_F\tau})$, where $\nu_F=m/2\pi$ is the single-particle density of states per spin. 

The computation of the current density requires the knowledge of the electronic distribution function. 
Here, following the avenue of Ref. \cite{Andrey2006}, I will use the Wigner distribution function (WDF) which is a $2\times 2$ matrix in spin space and is defined according to 
\beg\label{WDF}
\begin{split}
\hat{w}_{\bk\eps}(\br,t)&=\frac{1}{2\pi}\int d^2{\mathbf s}\int d\tau e^{i\bk{\mathbf s}-i\eps\tau}\\&\times\left\langle\psi_\beta\dg\left(\br+\frac{\mathbf s}{2},t+\frac{\tau}{2}\right)\psi_\alpha\left(\br-\frac{\mathbf s}{2},t-\frac{\tau}{2}\right)\right\rangle.
\end{split}
\en
Using the Dyson equations for the Keldysh function along with the definition (\ref{WDF}) one can show that the Wigner distribution function satisfies the following kinetic equation \cite{Supplementary}:
\begin{widetext}
\beg\label{Eq4w}
\begin{split}
&\left(\partial_t+\frac{\bk\cdot{\mbox{\boldmath $\nabla$}}}{m}+\frac{1}{\tau}\right)\hat{w}_{\bk\eps}+i\alpha_{\textrm{so}}[(\bk\times{\vec e}_z)\cdot{\mbox{\boldmath $\sigma$}},\hat{w}_{\bk \eps}]
=-\frac{1}{2}\left\{\frac{\bk}{m}+\alpha_{\textrm{so}}\hat{\mbox{\boldmath $\eta$}},\overline{\mbox{\boldmath $\nabla$}}\hat{w}_{\bk \eps}\right\}-\frac{e\alpha_{\textrm{so}}}{2\omega}\left[\hat{\mbox{\boldmath $\eta$}},\hat{\gamma}_{\omega}(\bk\eps;\br t)\right]\delta{\mathbf E}(\br,t)\\&-\frac{\alpha_{\textrm{so}}}{2}\left\{\hat{\mbox{\boldmath $\eta$}},{\mbox{\boldmath $\nabla$}}\hat{w}_{\bk \eps}\right\}
+\frac{i}{2\pi\nu_F\tau}\int\frac{d^2\bk}{(2\pi)^2}\left[\hat{G}_{\bk\eps}^R(\br,t)\circ\hat{w}_{\bk\eps}(\br,t)-\hat{w}_{\bk\eps}(\br,t)\circ\hat{G}_{\bk\eps}^A(\br,t)\right],
\end{split}
\en
\end{widetext}
where $\delta{\mathbf E}(\br,t)={\mathbf E}_0e^{i(\bq\br-\omega t)}-{\mathbf E}_0^*e^{-i(\bq\br-\omega t)}$, $\hat{G}_{\bk\eps}^{R(A)}(\br,t)$ are retarded (advanced) functions and
\beg\label{overn}
\overline{\mbox{\boldmath $\nabla$}}\hat{w}_{\bk\veps}=\frac{e}{\omega}\left(\hat{w}_{\bk \veps+\frac{\omega}{2}}-\hat{w}_{\bk \veps-\frac{\omega}{2}}\right){\mathbf E}(\br,t).
\en
It is easy to see that in limit of small frequencies the right hand side of the expression (\ref{overn}) acquires a familiar form $e{\mathbf E}\partial_\eps\hat{w}_{\bk\veps}$ \cite{MSH2004}. Notably, the appearance of an additional term proportional to $\delta{\mathbf E}(\br,t)$ and which has never been discussed in the literature before, is somewhat unusual. Upon closer examination one finds that function $\hat{\gamma}_{\omega}(\bk\eps;\br t)$ vanishes in the dc limit since the first non-zero contribution to this function is proportional to $\omega^2$ at small frequencies. One therefore can expect that the terms proportional to $\hat{\gamma}_{\omega}(\bk\eps;\br t)$ should only contribute to transport in the nonlinear regime. The expression for the functions $\hat{\gamma}_{\omega}(\bk\eps;\br t)$ in terms of the WDF can be found in the Supplementary Materials. It is also shown there that as far as the IFE is concerned the contributions to the current density which involve $\hat{\gamma}_{\omega}(\bk\eps;\br t)$ are at least a factor of $(\omega/\veps_F)^2$ smaller than the leading one \cite{Supplementary}. Lastly, in equation (\ref{Eq4w}) the convolution should be understood in the usual sense as
$A_{\bp\veps}(t)\circ B_{\bp\veps}(t)=A_{\bp\veps}(t)e^{\frac{i}{2}(\stackrel{\leftarrow}\partial_\veps\stackrel{\rightarrow}\partial_t-\stackrel{\leftarrow}\partial_t\stackrel{\rightarrow}\partial_\veps)}B_{\bp\veps}(t)$. Equation (\ref{Eq4w}) includes the spatial and time derivatives as well as dependence on relative momentum and energy. As such it constitutes to a quantum kinetic equation for the Wigner distribution function.

It is straightforward to check that in equilibrium, when spin-orbit coupling is zero for the WDF, one finds $\tilde{w}_{\alpha\beta}^{(0)}(\bk,\eps)=f(\eps)\delta(\eps-\eps_k)\delta_{\alpha\beta}$,
where $\eps_k$ denotes the single particle energy, $f(\eps)=\{\exp[(\eps-\mu)/k_BT]+1\}^{-1}$ is the Fermi-Dirac distribution function, $T$ is temperature and $\mu$ is the chemical potential. Consequently, in the presence of the spin-orbit coupling the WDF acquires the off-diagonal spin components which account for the possibility of having momentum pointing along or opposite to the direction of the electron's spin
$\hat{w}_{\bk\eps}^{(0)}=(1/2)\sum_{s=\pm}\left(\delta_{\alpha\beta}+s\bn\cdot\hat{\mbox{\boldmath $\eta$}}\right)
f(\eps)\delta\left(\eps-E_{\bk s}\right)$
and $\bn=\bk/k$. It is worth to point out here that the last term on the right hand side of (\ref{Eq4w}) vanishes in equilibrium. 

\paragraph{Current density.}
The charge current density is defined by (see e.g. \cite{Andrey2006}):
\beg\label{current}
{\mathbf j}(\br,t)=\frac{e}{m}\int\limits_0^\infty\frac{k^2dk}{2\pi}\int\limits_0^{2\pi}\frac{d\theta}{2\pi}\int\limits_{-\infty}^{\infty}\bn \textrm{Tr}[\hat{w}_{\bk\eps}(\br,t)]d\eps 
\en
In the context of the inverse Faraday effect, the problem consists in separating from all the possible contributions to the Wigner distribution function only those for which the corresponding corrections to the current density will contain the following vector combination $\hat{w}_{\bk\eps}^{(2)}\propto{\mathbf E}_0(\bq{\mathbf E}_0^*)+(\bq{\mathbf E}_0){\mathbf E}_0^*\to i{\mbox{\boldmath $\nabla$}}\times({\mathbf E}\times{\mathbf E}^*)$ \cite{dzero2024ife,Balatsky2023}.  In the Supplementary Materials I provide the details on the perturbative solution of the kinetic equation (\ref{Eq4w}) under the assumption that the electric field is weak. 

Since there is no current flowing in the ground state, the first nonzero contribution to the current is linearly proportional to electric field. After computing the trace over the spin indices, Eq. (\ref{current}), the linear-in-field expression for the current density reads \cite{Supplementary}:
\beg\label{current1}
\begin{split}
{\mathbf j}^{(1)}(\br,t)&=\frac{e^2\tau}{m\omega}e^{i(\bq\br-\omega t)}\sum\limits_{s=\pm}\int\limits_{-\infty}^{\infty}d\eps\int\limits_0^{2\pi}\frac{d\theta}{2\pi}\bn(\bn{\mathbf E}_0)\\&\times
\int\limits_0^\infty\frac{k^2dk}{2\pi}\frac{(k-sm\alpha_{\textrm{so}})}{(1-i\omega\tau)}\left(F_{\bk\eps-\frac{\omega}{2}}^{(s)}-F_{\bk\eps+\frac{\omega}{2}}^{(s)}\right),
\end{split}
\en
where I set $\bq=0$ in the expression under the integral since $v_Fq/\omega=v_F/c\ll1$ and  used the following notation
$F_{\bk\eps}^{(s)}=f(\eps)\delta\left(\eps-\eps_\bk-s\alpha_{\textrm{so}}k\right)$ \cite{Supplementary}.
It is easy to check that upon the integration over the remaining variables one recovers the Drude formula ${\mathbf j}^{(1)}=n_{\textrm{c}}e^2\tau{\mathbf E}/m(1-i\omega\tau)$, where $n_{\textrm{c}}$ is the carrier density. Note that the anomalous term which contains function 
$\hat{\gamma}_\omega$ does not enter into the final expression for ${\mathbf j}^{(1)}$, Eq. (\ref{current1}), as it was expected. 

\begin{figure}
\includegraphics[width=0.75\linewidth]{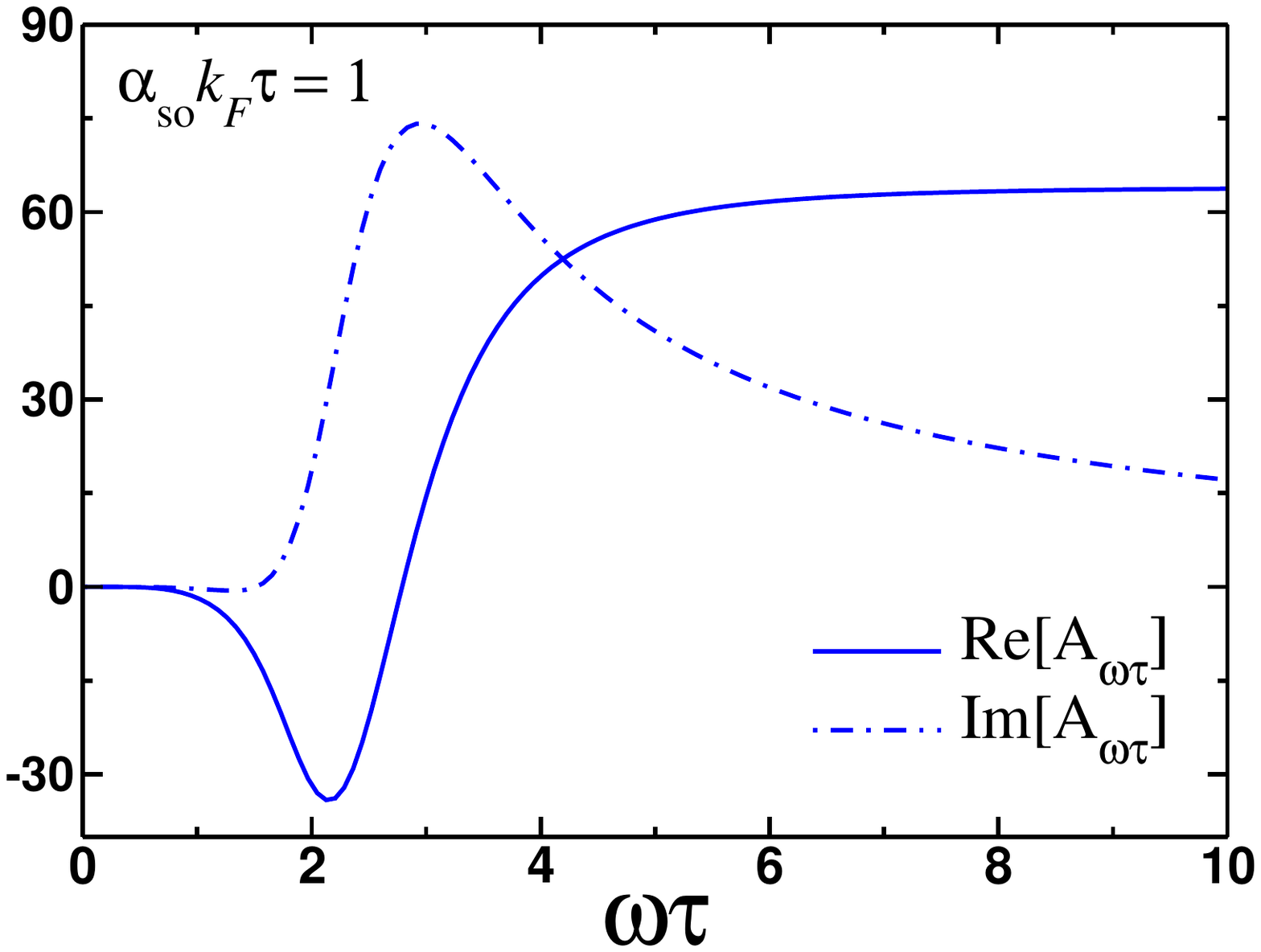}
\caption{The plot of the real and imaginary parts of $A_{\omega\tau}$, Eq. (\ref{Awt}),  as a function of $\omega\tau$. The value of the dimensionless parameter $\zeta=\alpha_{\textrm{so}}k_F\tau$ has been fixed to $\zeta=1$. The peak in $A_{\omega\tau}''$ and dip in $A_{\omega\tau}'$ at $\omega\tau\approx 2\zeta$ are governed by the same physical processes: interband transitions between two spin-orbit split electronic bands. I also note that function $A_{\omega\tau}'$ which determines the magnitude of the effect changes sign depending on the value of $\omega\tau$.}
\label{FigAwtau}
\end{figure}

Let me now discuss the second order corrections to the current density. As before, it is convenient to compute the trace over the spin indices. As a result I find five distinct contributions: the first two contain the vector combinations of the form $2n_xn_y(E_y^*E_x+E_x^*E_y)$
and $(n_x^2-n_y^2)(E_x^*E_x-E_y^*E_y)$ which clearly vanish due imposed condition that ${\mathbf E}({\mathbf r},t)$ must be  circularly polarized, while the third contribution is proportional to scalar product $({\mathbf E}_0\cdot{\mathbf E}_0^*)$ and as such will be of no interest to the present discussion \cite{Edel1988}. Thus, only two remaining contributions are relevant for the inverse Faraday effect. The intermediate expressions which arise during the calculation are listed in the Supplementary Materials and here I will list the final results only. Specifically,  the first from the remaining two contributions to $\textrm{Tr}[\hat{w}_{\bk\eps}(\br,t)]$ results in following expression for the current density 
\beg\label{j1}
\begin{split}
{\mathbf j}_{\textrm{IFE}}^{({\textrm{2})}}&=\frac{3m}{2\pi}\left(\frac{e\alpha_{\textrm{so}}}{\omega}\right)^3\left(\frac{\veps_F}{\omega}\right)^2\frac{A_{\omega\tau}}{m\alpha_{\textrm{so}}}\left\{{\mbox{\boldmath $\nabla$}}\times\left(i\eps_0{\mathbf E}\times{\mathbf E}^*\right)\right\},
\end{split}
\en
where $\eps_0$ is a permittivity in vacuum and I introduced a complex valued dimensionless function $A_{\omega\tau}=A_{\omega\tau}'+iA_{\omega\tau}''$:
\beg\label{Awt}
A_{\omega\tau}=-\frac{(2\omega\tau)^5}{[1+(\omega\tau)^2]^2}\frac{[4\omega\tau\zeta^2+i(1-2i\omega\tau)(1+i\omega\tau)^2]}{[(1+i\omega\tau)^2+4\zeta^2]^2}.
\en
It is worth noting that the expression for the current density (\ref{j1}) is somewhat similar to the one found for the second harmonic of the current density when static magnetic field is applied parallel to the plane of motion \cite{Edel1988}. In the derivation of the expression (\ref{j1}) I took into account that the main contribution to the integral over momentum in (\ref{current}) comes from the region $k\sim k_F$. In the limit of large frequencies and small relaxation rate ($\omega\tau\gg 1$) function $A_{\omega\tau\gg 1}'$ remains positive and $A_{\omega\tau\gg 1}'\approx64$. In the opposite limit of small frequency, $\omega\tau\ll1$,
function $A_{\omega\tau}'\sim-[(2\omega\tau)^6/(1+4\zeta^2)](\tau/\tau_s)$ acquires negative values. In Fig. \ref{FigAwtau} I present the dependence of $A_{\omega\tau}'$ and $A_{\omega\tau}''$ is a function of $\omega\tau$ which indeed shows that both $A_{\omega\tau}'$ and $A_{\omega\tau}''$ are non-monotonic functions of frequency exhibiting an extremum at frequencies $\omega^*\approx2\alpha_{\textrm{so}}k_F$. The appearance of extrema has a clear physical interpretation: it describes the contributions from the electronic transitions between two spin-orbit split bands.
Lastly, I note that under assumption that $\omega\sim \alpha_{\textrm{so}}k_F$ the dimensionless ratio $q/(m\alpha_{\textrm{so}})$ is of the order of $v_F/c\ll 1$, the large pre-factor $(\veps_F/\omega)^2$ in (\ref{j1}) guarantees that the value of the induced magnetization will be large enough to be detected experimentally.  

As for another contribution to the current density, upon closer examination it turns out it is of the order of $\omega/\veps_F$ smaller than ${\mathbf j}_{\textrm{IFE}}^{({\textrm{2})}}$, Eq. (\ref{j1}) \cite{Supplementary}. Thus, given the expression ${\mathbf j}_{\textrm{IFE}}^{({\textrm{2})}}={\mbox{\boldmath $\nabla$}}\times{\mathbf M}_{\textrm{ind}}$, the frequency dependence of the induced static magnetization will be approximately given by the following formula: 
\beg\label{Mind}
{\mathbf M}_{\textrm{ind}}(\omega)\approx A_{\omega\tau}'\frac{3e^3}{2\pi}\frac{(\alpha_{\textrm{so}}\veps_F)^2}{\omega^5}
\left(i\eps_0{\mathbf E}\times{\mathbf E}^*\right).
\en 

Expressions (\ref{Awt},\ref{Mind}) are the main results of this paper. The striking feature of these results is that in accordance with the discussion above it shows that the direction of the induced magnetization may change depending on the value of $\omega$, disorder scattering rate $\tau^{-1}$ and parameter $\zeta$. Since non-monotonic dependence of $A_{\omega\tau}$ is due to the optical transitions between the two spin-orbit split bands, the switching direction of the magnetization must be governed by the processes pertaining to the existence of the two bands with different chirality. In particular, the frequency when the magnetization vanishes must approximately correspond to the regime when the kinematic processes in the two chiral bands which contribute to the IFE cancel each other. For this reason, this result should hold independently of particular choice of the momentum dependence of the spin-orbit coupling or type of the disorder scattering. This finding therefore opens an exciting new way for manipulating the emerging magnetization by varying the frequency of an external light. 

\paragraph{Discussion.}
The frequency dependence of the magnetization, Eq. (\ref{Mind}), is quite different from the expected $1/\omega^3$ frequency dependence reported in previous studies \cite{Edelstein1998,Balatsky2023,dzero2024ife}. It must be emphasized here that strictly speaking the derivation presented above along with the result for the magnetization is certainly applicable when the disorder scattering rate is small since the contributions from the disorder collision integral - defined by the last term on the right hand side of Eq. (\ref{Eq4w}) - has been neglected. In the opposite case of strong impurity scattering the frequency dependence $M_{\textrm{ind}}\propto \omega^{-5}$ may change at least in the limit of high frequencies $\omega\tau\gg1$  to $M_{\textrm{ind}}(\omega)\propto \omega^{-3}$ due to the contribution from the collision integral term that have been neglected. Indeed, Ref. \onlinecite{Edelstein1998} reports for magnetization $M_{\textrm{ind}}\propto 1/(\omega\tau)^{3}$ which clearly vanishes in the clean limit, but may become dominant in the dirty limit. Specifically, in the dirty limit, defined by $\zeta\ll 1$,  it is in principle possible that the strong impurity scattering described by the collision integral may introduce the cutoffs such that two powers of frequency will be replaced with the square of the disorder scattering rate. In other words, the quantum kinetic equation without disorder induced collision integral part only accounts for scattering processes which correspond to  effects of disorder in equilibrium as well as out-of-equilibrium, while the collision integral involves disorder scattering out-of-equilibrium only. For this reason, one needs to keep in mind that the main result, Eq. (\ref{Mind}), may also be applicable in the case when disorder scattering rate is small, $\veps_F\tau\gg 1$. In this case the corresponding corrections to magnetization can be taken into account by perturbation theory. 

The model that has been used here is special in two aspects: the spin-orbit coupling term in the model Hamiltonian (\ref{Eq1}) has a linear momentum dependence and the disorder potential is assumed to be local. The same model has been used, for example, in the analysis of the spin-Hall effect  \cite{MSH2004} where it was shown that the spin Hall conductivity vanishes in the dc limit. At the same time, generalization of this model to the case of short-range disorder potential and generic momentum dependence of the spin-orbit coupling yields finite value of the dc spin-Hall conductivity, which can acquire either positive or negative values \cite{Andrey2006}. Thus, in principle, one may expect that the frequency dependence found here is not universal and may in fact change with the variation in the nature of the disorder potential and the details of the spin-orbit coupling. However, as it has been already mentioned above, the effect of magnetization switching its direction depending on external frequency, should not depend on the details of the underlying microscopic model and, as such, should be universal. 

On a more general note, it is clear that while the idea of the inverse Faraday effect being induced by the spin-orbit coupling is certainly a correct one, the treatment proposed in Refs. \cite{Edelstein1998,TaguchiTatara2011} has several important shortcomings. Not only the derivations are based on using the diagrammatic technique in equilibrium, while the effect is manifestly non-equilibrium one, but they also miss the contributions to magnetization from several important processes such as interband transitions. These processes are, of course, necessarily present in the second order corrections to the current density. Whether the discrepancies between my results and those of Refs. \cite{Edelstein1998,TaguchiTatara2011} are due to those theoretical shortcomings or not still remains to be better understood. I leave the problems outlined above for future studies. 

\paragraph{Conclusions.}
In this paper I have used the methods of non-equilibrium quantum field theory to provide a systematic derivation of the quantum kinetic equation for the Wigner distribution function, which describes the dynamics of the two-dimensional disordered metallic system with broken mirror symmetry. The non-equilibrium conditions are assumed to be induced by circularly polarized monochromatic light. The main goals of this study were to calculate the second order corrections to the current density in powers of the electric field amplitude and elucidate the nonlocal contributions to the current density which describe the emergence of the static magnetization. Assuming that disorder scattering is weak,  I found that the magnitude of the effect is proportional to the square of the spin-orbit coupling and at high frequencies is inversely proportional to 
fifth power of frequency. In the opposite limit of small frequencies the magnetization will be proportional to the frequency of an external electromagnetic wave. Importantly, I found that the direction of the induced magnetization also depends on the frequency of light. This effect may be used for implementing yet another mechanism for the ultrafast manipulation of magnetic states by external electromagnetic radiation in novel materials. 

\paragraph{Acknowledgments.}
This work was financially supported by the National Science Foundation grant DMR-2002795. Significant part of this work has been carried out during the Aspen Center for Physics 2024 Summer Program on "\emph{Probing Collective Excitations in Quantum Matter by Transport and Spectroscopy}", which was supported by the National Science Foundation Grant No. PHY-2210452. 


\begin{thebibliography}{10}

\bibitem{IFE-Exp-1965}
J.~P. van~der Ziel, P.~S. Pershan, and L.~D. Malmstrom, ``Optically-induced
  magnetization resulting from the inverse Faraday effect,'' {\em Phys. Rev.
  Lett.}, vol.~15, pp.~190--193, Aug 1965.

\bibitem{Pit1961}
L.~P. Pitaevskii, ``Electric forces in a transparent dispersive medium,'' {\em
  Sov. Phys. - JETP}, vol.~12, p.~1008, 1961.

\bibitem{Pershan1966}
P.~S. Pershan, J.~P. van~der Ziel, and L.~D. Malmstrom, ``Theoretical
  discussion of the inverse Faraday effect, Raman scattering, and related
  phenomena,'' {\em Phys. Rev.}, vol.~143, pp.~574--583, Mar 1966.

\bibitem{Edelstein1998}
V.~M. Edelstein, ``Inverse Faraday effect in conducting crystals caused by a
  broken mirror symmetry,'' {\em Phys. Rev. Lett.}, vol.~80, pp.~5766--5769,
  Jun 1998.

\bibitem{Battiato2014}
M.~Battiato, G.~Barbalinardo, and P.~M. Oppeneer, ``Quantum theory of the
  inverse Faraday effect,'' {\em Phys. Rev. B}, vol.~89, p.~014413, Jan 2014.

\bibitem{yang2022inverse}
X.~Yang, Y.~Mou, H.~Zapata, B.~Reynier, B.~Gallas, and M.~Mivelle, ``An inverse
  Faraday effect through linear polarized light,'' 2022.

\bibitem{mou2023reversed}
Y.~Mou, X.~Yang, B.~Gallas, and M.~Mivelle, ``A reversed inverse Faraday
  effect,'' 2023.

\bibitem{mou2023chiral}
Y.~Mou, X.~Yang, B.~Gallas, and M.~Mivelle, ``A chiral inverse Faraday effect
  mediated by an inversely designed plasmonic antenna,'' 2023.

\bibitem{Putilov2023-IFESC}
A.~V. Putilov, S.~V. Mironov, A.~S. Mel'nikov, and A.~A. Bespalov, ``Inverse
  Faraday effect in superconductors with a finite gap in the excitation
  spectrum,'' {\em JETP Letters}, vol.~117, no.~11, pp.~827--833, 2023.

\bibitem{Mironov2021-IFESC}
S.~V. Mironov, A.~S. Mel'nikov, I.~D. Tokman, V.~Vadimov, B.~Lounis, and A.~I.
  Buzdin, ``Inverse Faraday effect for superconducting condensates,'' {\em
  Phys. Rev. Lett.}, vol.~126, p.~137002, Apr 2021.

\bibitem{Buzdin2023}
V.~Plastovets and A.~Buzdin, ``Fluctuation-mediated inverse Faraday effect in
  superconducting rings,'' {\em Physics Letters A}, vol.~481, p.~129001, 2023.

\bibitem{Parafilo2022Fl}
A.~V. Parafilo, M.~V. Boev, V.~M. Kovalev, and I.~G. Savenko, ``Photogalvanic
  transport in fluctuating Ising superconductors,'' {\em Phys. Rev. B},
  vol.~106, p.~144502, Oct 2022.

\bibitem{Croitoru2023}
M.~D. Croitoru, B.~Lounis, and A.~I. Buzdin, ``{Helicity-controlled switching
  of superconducting states by radiation pulse},'' {\em Applied Physics
  Letters}, vol.~123, p.~122601, 09 2023.

\bibitem{IFE-Mott2022}
S.~Banerjee, U.~Kumar, and S.-Z. Lin, ``Inverse Faraday effect in Mott
  insulators,'' {\em Phys. Rev. B}, vol.~105, p.~L180414, May 2022.

\bibitem{gao2020topological}
Y.~Gao, C.~Wang, and D.~Xiao, ``Topological inverse Faraday effect in Weyl
  semimetals,'' 2020.

\bibitem{parchenko2023plasmonenhanced}
S.~Parchenko, K.~Hofhuis, A.~Ciuciulkaite, V.~Kapaklis, V.~Scagnoli,
  L.~Heyderman, and A.~Kleibert, ``Plasmon-enhanced optical control of
  magnetism at the nanoscale via the inverse Faraday effect,'' 2023.

\bibitem{Han_2023}
J.~W. Han, P.~Sai, D.~B. But, E.~Uykur, S.~Winnerl, G.~Kumar, M.~L. Chin, R.~L.
  Myers-Ward, M.~T. Dejarld, K.~M. Daniels, T.~E. Murphy, W.~Knap, and
  M.~Mittendorff, ``Strong transient magnetic fields induced by THz-driven
  plasmons in graphene disks,'' {\em Nature Communications}, vol.~14, Nov.
  2023.

\bibitem{Balatsky2023}
P.~Sharma and A.~V. Balatsky, ``Light-induced orbital magnetism in metals via
  inverse Faraday effect,'' {\em Phys. Rev. B}, vol.~110, p.~094302, Sep 2024.

\bibitem{dzero2024ife}
M.~Dzero, ``Inverse Faraday effect in superconductors with potential  impurities,'' {\em Phys. Rev. B}, vol.~110, p.~054506, Aug 2024.

\bibitem{Kulik1981}
I.~O. Kulik, O.~Entin-Wohlman, and R.~Orbach, ``Pair susceptibility and mode
  propagation in superconductors: A microscopic approach,'' {\em Journal of Low
  Temperature Physics}, vol.~43, pp.~591--620, Jun 1981.


\bibitem{TaguchiTatara2011}
K.~Taguchi and G.~Tatara, ``Theory of inverse Faraday effect in a disordered
  metal in the terahertz regime,'' {\em Phys. Rev. B}, vol.~84, p.~174433, Nov
  2011.

\bibitem{Juraschek2017}
D.~M. Juraschek, M.~Fechner, A.~V. Balatsky, and N.~A. Spaldin, ``Dynamical
  multiferroicity,'' {\em Phys. Rev. Mater.}, vol.~1, p.~014401, Jun 2017.

\bibitem{AxialMagneto2021}
L.~Liang, P.~O. Sukhachov, and A.~V. Balatsky, ``Axial magneto-electric effect
  in Dirac semimetals,'' {\em Phys. Rev. Lett.}, vol.~126, p.~247202, Jun 2021.

\bibitem{MSH2004}
E.~G. Mishchenko, A.~V. Shytov, and B.~I. Halperin, ``Spin current and
  polarization in impure two-dimensional electron systems with spin-orbit
  coupling,'' {\em Phys. Rev. Lett.}, vol.~93, p.~226602, Nov 2004.

\bibitem{Andrey2006}
A.~V. Shytov, E.~G. Mishchenko, H.-A. Engel, and B.~I. Halperin, ``Small-angle
  impurity scattering and the spin-Hall conductivity in two-dimensional
  semiconductor systems,'' {\em Phys. Rev. B}, vol.~73, p.~075316, Feb 2006.

\bibitem{Kamenev2009}
A.~Kamenev and A.~Levchenko, ``Keldysh technique and non-linear $\sigma$-model:
  basic principles and applications,'' {\em Advances in Physics}, vol.~58,
  no.~3, pp.~197--319, 2009.

\bibitem{Kamenev2011}
A.~Kamenev, {\em Field Theory of Non-Equilibrium Systems}.
\newblock Cambridge University Press, 2011.

\bibitem{Edel1988}
V.~M. Edelstein, ``Second-harmonic generation in two-dimensional systems
  without inversion centers,'' {\em Sov. Phys. - JETP}, vol.~68, p.~1446, 1988.

\bibitem{Kita-Review}
T.~Kita, ``{Introduction to Nonequilibrium Statistical Mechanics with Quantum
  Field Theory},'' {\em Progress of Theoretical Physics}, vol.~123,
  pp.~581--658, 04 2010.

\bibitem{Kita2001}
T.~Kita, ``Gauge invariance and Hall terms in the quasi-classical equations of
  superconductivity,'' {\em Phys. Rev. B}, vol.~64, p.~054503, Jun 2001.

\bibitem{VK1973}
A.~F. Volkov and S.~M. Kogan, ``Collisionless relaxation of the energy gap in
  superconductors,'' {\em Zh. Eksp. Teor. Fiz}, vol.~65, p.~2038, 1974.
\newblock {English} translation: Sov. Phys. JETP, {\bf 38}, 1018 (1974).

\bibitem{Supplementary}
See Supplementary Materials for details.

\end{thebibliography}

\clearpage

\begin{appendix}
\begin{widetext}
\begin{center}
{\large\bf Supplementary Materials}
\end{center}
\vspace{0.1cm}

Here I provide additional technical details on the derivation of the main equations listed in the main text.

\section{Quasiclassical equation for the gauge-invariant Keldysh function}\label{AppEq4GK}
In this Section of the Supplementary Materials I derive the quasiclassical equation for the Keldysh function, which is later used to derive the quantum kinetic equation for the Wigner distribution function.

My goal here is to compute the current density induced by external field in the system governed by the model Hamiltonian (\ref{Eq1}) in the main text. Since the phenomenon that I consider takes place when a system is driven out-of-equilibrium, for the purposes of computing the induced current I introduce the retarded $\hat{G}^R$, advanced $\hat{G}^A$ and Keldysh $\hat{G}^K$ Green's functions which are $2\times 2$ matrices in the spin space and they satisfy the Dyson equations \cite{Kamenev2009,Kamenev2011}
\beg\label{Eq2}
\left(\hat{\cal G}_0^{-1}-\check{\Sigma}\right)\circ\check{G}=\check{1}, \quad \check{G}\circ\left(\hat{\cal G}_0^{-1}-\check{\Sigma}\right)=\check{1},
\en
where 
\beg\label{CheckG}
\check{G}=\left(\begin{matrix} \hat{G}^R & \hat{G}^K \\ 0 & \hat{G}^A \end{matrix}\right)
\en
and
$\hat{\cal G}_0^{-1}=i\partial_{t}-\hat{\cal H}(\br,t)$. The self-energy part appearing in (\ref{Eq2}) is determined within the self-consistent Born approximation to be equal to $\check{\Sigma}(x,x')={\delta(\br-\br')}\check{G}(x,x')/{2\pi\nu_F\tau}$,  
and I used the shorthand notation $x=(\br,t)$ \cite{MSH2004}.

Given the form of the Hamiltonian (see Eq. (\ref{Eq1}) in the main text) after performing the Fourier transform for the retarded and advanced functions in the ground state I find \cite{Edel1988}
\beg\label{GRGA}
\hat{G}_{\bp\eps}^{R(A)}=\frac{1}{2}\left(\frac{1+(\bp/p)\cdot\hat{\mbox{\boldmath $\eta$}}}{\eps-E_{\bp,+}\pm \frac{i}{2\tau}}+\frac{1-(\bp/p)\cdot\hat{\mbox{\boldmath $\eta$}}}{\eps-E_{\bp,-}\pm\frac{i}{2\tau}}\right).
\en
Here $E_{\bp,\pm}=p^2/2m\pm \alpha_{\textrm{so}}p$ determines the single-particle spectrum of two spin-orbit split bands.

The information about the nonequilibrium state of a system is contained in the function $\hat{G}^K$. Using the Dyson equation (\ref{Eq2}) it is straightforward to show that the Keldysh component of $\check{G}(x,x')$ satisfies the relation 
\beg\label{GKGRGA}
\hat{G}^K=\hat{G}^R\circ\hat{\Sigma}^K\circ\hat{G}^A.
\en
From this relation we derive the following equation for the Keldysh function \cite{MSH2004,Andrey2006}:
\beg\label{Eq4GK}
\left[\hat{G}^R\right]^{-1}\circ\hat{G}^K-\hat{G}^K\circ\left[\hat{G}^A\right]^{-1}=\hat{\Sigma}^K\circ\hat{G}^A-\hat{G}^R\circ\hat{\Sigma}^K.
\en
Using equation (\ref{Eq4GK}), one performs the Wigner transformation of the Keldysh function with respect to the relative coordinates $x-x'$ and obtains the quasiclassical equation for the corresponding Wigner-transformed Keldysh function \cite{Kamenev2009,Kamenev2011}. Furthermore, it is also reasonable to work with the gauge-invariant functions \cite{Kita-Review,Kita2001,dzero2024ife,Balatsky2023}.

 I now apply the Wigner transformation to the both sides of equation (\ref{Eq4GK}):
\beg\label{Wigner}
\begin{split}
&\hat{G}^K(\br,t;\br',t')=\int\frac{d\eps}{2\pi}\int\frac{d^2\bk}{(2\pi)^2}e^{i\bk(\br-\br')-i\eps(t-t')}\hat{g}_{\bk \eps}\left(\frac{\br+\br'}{2},\frac{t+t'}{2}\right).
\end{split}
\en
For the expression on the left hand side (\ref{Eq4GK}) for the gauge invariant propagator I have
\beg\label{LHS}
\begin{split}
&[\hat{G}^R]^{-1}\circ\hat{G}^K-\hat{G}^K\circ[\hat{G}^A]^{-1}=i\int\frac{d\eps}{2\pi}\int\frac{d^2\bk}{(2\pi)^2}e^{i\bk(\br-\br')-i\eps(t-t')}\left[\frac{\partial \hat{g}_{\bk\eps}}{\partial t}+\frac{1}{2}\left\{\frac{\bk}{m}+\alpha_{\textrm{so}}\hat{\mbox{\boldmath $\eta$}},{\mbox{\boldmath $\nabla$}}\hat{g}_{\bk\eps}\right\}\right.\\&\left.+i\alpha_{\textrm{so}}[(\bk\times{\vec e}_z\cdot{\mbox{\boldmath $\sigma$}},\hat{g}_{\bk\eps}]+\frac{\partial \hat{g}_{\bk\eps}}{\partial\eps}\hat{\cal E}[(i/2)\stackrel{\leftarrow}\partial_\eps\stackrel{\rightarrow}\partial_t-(i/2)\stackrel{\leftarrow}\partial_\bk\stackrel{\rightarrow}\partial_\br]\left(\frac{\bk}{m}+\alpha_{\textrm{so}}\hat{\mbox{\boldmath $\eta$}}\right){\mathbf E}(\br,t)\right.
\\&\left.+\left(\frac{\bk}{m}+\alpha_{\textrm{so}}\hat{\mbox{\boldmath $\eta$}}\right){\mathbf E}(\br,t)
\hat{\cal E}[(i/2)\stackrel{\leftarrow}\partial_\br\stackrel{\rightarrow}\partial_\bk-(i/2)\stackrel{\leftarrow}\partial_t\stackrel{\rightarrow}\partial_\eps]\frac{\partial \hat{g}_{\bk \eps}}{\partial \eps}\right]+\hat{G}^K\circ\hat{\Sigma}^A-
\hat{\Sigma}^R\circ\hat{G}^K.
\end{split}
\en

In order to derive the kinetic equation for the WDF it is useful to recall the definition of the Green's functions on the Keldysh contour \cite{VK1973}
\beg\label{Gij}
{G}_{\alpha\beta}^{(kl)}(1,1')=\frac{1}{i}\langle \hat{T}_t\psi_\alpha(1_k)\psi_\beta\dg(1_l')\rangle, \quad (k,l=1,2).
\en
Here $'1'$ is a short notation for space and time coordinates, $l=1$ refers to the upper branch of the Keldysh contour which runs from $-\infty$ to $+\infty$, while $l=2$ refers to the lower branch of the Keldysh contour which runs from $+\infty$ to $-\infty$. From this definition it follows that WDF can be expressed in terms of the 'lesser' correlation function
\beg\label{WDFG12}
\hat{w}_{\bk\eps}(\br,t)=\left(-\frac{i}{2\pi}\right)\hat{G}_{\bk\eps}^{(12)}(\br,t).
\en
Given the relation 
\beg\label{G12}
\hat{G}_{\bk\eps}^{(12)}(\br,t)=\frac{1}{2}\left(\hat{G}_{\bk\eps}^{K}(\br,t)-\hat{G}_{\bk\eps}^{R}(\br,t)+\hat{G}_{\bk\eps}^{A}(\br,t)\right),
\en
one can employ (\ref{WDFG12}) along with the expressions (\ref{GRGA}) for the retarded and advanced functions to reproduce expression for the distribution function $\hat{w}_{\bk\eps}^{(0)}$ in the main text.
One can also immediately realize that it is possible write down an equation for the Wigner distribution function using equations for the retarded and advanced functions which follow from the Dyson equations (\ref{Eq2}) and equation for the Keldysh function.

On the right hand side of this expression 
operator $\hat{\mbox{\boldmath $\eta$}}$ is defined according to $\hat{\mbox{\boldmath $\eta$}}={\vec e}_z\times\hat{\mbox{\boldmath $\sigma$}}$ In view of the fact $q={\omega}/{c}\ll {\omega}/{v_F}$
I will be neglecting the derivatives with respect to $\bk$ and $\br$ in the arguments of operator functions $
\hat{\cal E}$ in expression (\ref{LHS}).
\section{Operator function ${\cal E}(\hat{u})$}
I will ignore the spatial derivatives in the arguments of function $\hat{\cal E}(\hat{u})$ which is defined as follows:
\beg\label{Eu}
{\cal E}(\hat{u})=\frac{e}{4}\left[2{\cal E}_1(\hat{u})-{\cal E}_2(\hat{u})+{\cal E}_2(-\hat{u})\right],
\en
where 
${\cal E}_1(\hat{u})=({e^{\hat{u}}-1})/{\hat{u}}$ and ${\cal E}_2(\hat{u})=({e^{\hat{u}}-\hat{u}-1)}/{\hat{u}^2}$. It obtains:
\beg\label{EuAgain}
\begin{split}
&{\cal E}(\hat{u})=\frac{e}{4}\left[{\cal E}_1(\hat{u})+{\cal E}_1\left(-\hat{u}\right)+\int\limits_0^1 \eta e^{\eta\hat{u}}d\eta-\int\limits_0^1 \eta e^{-\eta\hat{u}}d\eta\right].
\end{split}
\en
Note that the last term in this expression is an odd function of $\hat{u}$. 
Taking into account expression
\beg\label{Derivs}
\begin{split}
&\frac{e}{4}\frac{\partial \hat{g}_{\bk\veps}}{\partial\veps}\left[{\cal E}_1\left(\frac{i}{2}\stackrel{\leftarrow}\partial_t\stackrel{\rightarrow}\partial_\veps\right)+{\cal E}_1\left(-\frac{i}{2}\stackrel{\leftarrow}\partial_t\stackrel{\rightarrow}\partial_\veps\right)\right]{\mathbf E}_+(\br,t)=\frac{e}{2\omega}\left(
\hat{g}_{\bk{\veps+\frac{\omega}{2}}}-\hat{g}_{\bk{\veps-\frac{\omega}{2}}}\right){\mathbf E}_+(\br,t).
\end{split}
\en
I have
\beg\label{FirstPart}
\begin{split}
&\frac{\partial \hat{g}_{\bk\veps}}{\partial \veps}\hat{\cal E}\left(\frac{i}{2}\stackrel{\leftarrow}\partial_\veps\stackrel{\rightarrow}\partial_t\right)\frac{\bk}{m}{\mathbf E}_{+}+\frac{\bk}{m}{\mathbf E}_{+}
\hat{\cal E}\left(-\frac{i}{2}\stackrel{\leftarrow}\partial_t\stackrel{\rightarrow}\partial_\veps\right)
\frac{\partial \hat{g}_{\bk \veps}}{\partial\veps}=\frac{e}{m\omega}\left(
\hat{g}_{\bk{\veps+\frac{\omega}{2}}}-\hat{g}_{\bk{\veps-\frac{\omega}{2}}}\right)\bk{\mathbf E}_+,
\end{split}
\en
where I suppressed the arguments $(\br,t)$ of ${\mathbf E}_{+}(\br,t)={\mathbf E}_0e^{i(\bq\br-\omega t)}$ for brevity and the contributions from the last terms in (\ref{EuAgain}) cancel out. Notably, such a cancellation does not happen for the remaining term proportional to $\alpha_{\textrm{so}}$. Indeed, in this case one has
\beg\label{MyExtraTerm}
\begin{split}
&{\alpha_{\textrm{so}}}\frac{\partial \hat{g}_{\bk \veps}}{\partial \veps}\hat{\cal E}\left(\frac{i}{2}\stackrel{\leftarrow}\partial_\veps\stackrel{\rightarrow}\partial_t\right)\hat{\mbox{\boldmath $\eta$}}{\mathbf E}_{+}+
{\alpha_{\textrm{so}}}\hat{\mbox{\boldmath $\eta$}}{\mathbf E}_{+}
\hat{\cal E}\left(-\frac{i}{2}\stackrel{\leftarrow}\partial_t\stackrel{\rightarrow}\partial_\veps\right)
\frac{\partial \hat{g}_{\bk\veps}}{\partial\veps}=\frac{e\alpha_{\textrm{so}}}{2\omega}\left\{\hat{\mbox{\boldmath $\eta$}},\hat{g}_{\bk{\veps+\frac{\omega}{2}}}-\hat{g}_{\bk{\veps-\frac{\omega}{2}}}\right\}{\mathbf E}_{+}+
\frac{e\alpha_{\textrm{so}}}{2\omega}\left[\hat{\mbox{\boldmath $\eta$}},\hat{\Gamma}_{\bk\veps}(\omega)\right]{\mathbf E}_{+},
\end{split}
\en
where the dependence of ${\mathbf E}_{+}$ and $(\br,t)$ has been suppressed for brevity and I introduced function 
\beg\label{gammak}
\begin{split}
\hat{\Gamma}_{\bk\veps}(\omega)=\int\limits_0^1\left(\hat{g}_{\bk\veps+\eta\frac{\omega}{2}}+\hat{g}_{\bk\veps-\eta\frac{\omega}{2}}\right)d\eta-\hat{g}_{\bk\veps+\frac{\omega}{2}}-\hat{g}_{\bk\veps-\frac{\omega}{2}}.
\end{split}
\en
In the limit of small frequencies this function is manifestly proportional to $\omega^2$, which means that the last term in (\ref{MyExtraTerm}) will be nonzero only in the nonlinear ac regime. It is also clear that this term contributes to the second and higher harmonics of the current density. 

\section{Perturbative solution of the kinetic equation.}
To determine the linear in electric field correction to the WDF, I will look for the solution of the quantum kinetic equation (\ref{Eq4w}) in the form
\beg\label{Formw1}
\hat{w}_{\bk\veps}^{(1)}(\br,t)=\hat{w}_{\bk\veps}^{(1,+)}e^{i(\bq\br-\omega t)}+\hat{w}_{\bk\veps}^{(1,-)}e^{-i(\bq\br-\omega t)}.
\en
Before I discuss the details of the calculation, I am going to make further approximations which, as one can check, will not affect the subsequent results in any significant way. These approximations concern the last two terms which appear on the right hand side of the kinetic equation (\ref{Eq4w}). Specifically, assuming that the disorder scattering rate is sufficiently long, I will ignore the contributions to the WDF originating from the disorder collision integral. In addition, since in realistic situations $\alpha_{\textrm{so}}/v_F\ll1$ the corresponding contribution from the gradient term will also be ignored. In principle, the corrections originating from these two terms can be included at the later stages of the calculation.

In order to keep the subsequent formulas compact I represent the electric field similar to (\ref{Formw1}) as a sum of two counter propagating waves ${\mathbf E}_{+}(\br,t)$ and ${\mathbf E}_{-}(\br,t)$.
Keeping the terms of the first order in electric field, performing the Fourier transformation and using (\ref{Formw1}) I find
\beg\label{Eq4w1}
\begin{split}
&z_{qs}\hat{w}_{\bk\eps}^{(1,s)}+i\alpha_{\textrm{so}}[(\bn\times{\vec e}_z)\cdot{\mbox{\boldmath $\sigma$}},\hat{w}_{\bk \eps}^{(1,s)}]
=\hat{\cal K}^{(s)}[\hat{w}_{\bk\eps}^{(0)}],
\end{split}
\en
where $\hat{\eta}_\bk=({\vec e}_z\times{\mbox{\boldmath $\sigma$}})\cdot(\bk/k)$, $z_{qs}=-is(\omega-{\mathbf v}\cdot\bq)+\tau^{-1}$ and $s=\pm$ and I have introduced the kernel functions
\beg\label{K1pm}
\begin{split}
\hat{\cal K}^{(\pm)}[\hat{w}_{\bk\eps}^{(0)}]&=-\frac{e{\mathbf E}_\pm}{2\omega}\left\{\frac{\bk}{m}+\alpha_{\textrm{so}}\hat{\mbox{\boldmath $\eta$}},\hat{w}_{\bk \eps+\frac{\omega}{2}}^{(0)}-\hat{w}_{\bk \eps-\frac{\omega}{2}}^{(0)}\right\}\mp\frac{e\alpha_{\textrm{so}}}{2\omega}\left[\hat{\mbox{\boldmath $\eta$}}{\mathbf E}_{\pm},\hat{\gamma}_{\omega}^{(0)}(\bk,\eps)\right],
\end{split}
\en
Here I introduced function
\beg\label{SmallGamma}
\begin{split}
&\hat{\gamma}_{\omega}(\bk\eps;\br t)=\int\limits_0^1\left[\hat{w}_{\bk\veps+\eta\frac{\omega}{2}}(\br,t)+\hat{w}_{\bk\veps-\eta\frac{\omega}{2}}(\br,t)\right]d\eta-\hat{w}_{\bk\veps+\frac{\omega}{2}}(\br,t)-\hat{w}_{\bk\veps-\frac{\omega}{2}}(\br,t).
\end{split}
\en
Equation (\ref{Eq4w1}) can be formally solved.\cite{MSH2004}
In the matrix form of the solution reads
\beg\label{Formalw1}
\begin{split}
\hat{w}_{\bk\eps}^{(1,s)}&=\frac{(z_{qs}^2+2\Delta_k^2)}{z_{qs}(z_{qs}^2+4\Delta_k^2)}\hat{\cal K}^{(s)}+\frac{2\Delta_k^2}{z_{qs}(z_{qs}^2+4\Delta_k^2)}\hat{\eta}_\bk\hat{\cal K}^{(s)}\hat{\eta}_\bk-\frac{i\Delta_k}{z_{qs}^2+4\Delta_k^2}
\left[\hat{\eta}_\bk,\hat{\cal K}^{(s)}\right],
\end{split}
\en
where $\Delta_k=\alpha_{\textrm{so}}k$.

I continue with the calculation of the second order corrections. As one can easily check, there will be three terms contributing to the Wigner distribution function to the second order in electric field:
\beg\label{Formw2}
\hat{w}_{\bk\eps}^{(2)}(\br,t)=\hat{w}_{\bk\eps}^{(2,\textrm{dc})}+\hat{\tilde{w}}_{\bk\eps}^{(2,+)}e^{2i(\bq\br-\omega t)}+\hat{\tilde{w}}_{\bk\eps}^{(2,-)}e^{-2i(\bq\br-\omega t)}.
\en
In the context of the inverse Faraday effect I obviously need to determine $\hat{w}_{\bk\eps}^{(2,\textrm{d})}$. By construction $\hat{w}_{\bk\eps}^{(2,\textrm{dc})}$ is given by the sum of two contributions: 
$\hat{w}_{\bk\eps}^{(2,\textrm{dc})}=\hat{w}_{\bk\eps}^{(2,+)}+\hat{w}_{\bk\eps}^{(2,{-})}$ and 
functions $\hat{w}_{\bk\eps}^{(2,\pm)}$ are computed similarly to (\ref{Formalw1}):
\beg\label{wd2pm}
\begin{split}
\hat{w}_{\bk\eps}^{(2,s)}&=\frac{\tau(1+2\zeta_k^2)}{(1+4\zeta_k^2)}\hat{\cal K}_2^{(s)}+\frac{2\tau\zeta_k^2}{(1+4\zeta_k^2)}\hat{\eta}_\bk\hat{\cal K}_2^{(s)}\hat{\eta}_\bk-\frac{i\tau\zeta_k}{1+4\zeta_k^2}
\left[\hat{\eta}_\bk,\hat{\cal K}_2^{(s)}\right],
\end{split}
\en
where $\zeta_k=\tau\Delta_k$ and operator functions $\hat{\cal K}_2^{(\pm)}$ are defined according to $\hat{\cal K}_2^{(\pm)}=\hat{\cal K}^{(\pm)}[\hat{w}_{\bk\eps}^{(1)}]$. In passing I note that there will be no gradient correction to $\hat{w}_{\bk\eps}^{(2,\textrm{dc})}$ from the spin-orbit coupling term $\propto\hat{\gamma}_\omega$. Having expressed the second order corrections to the WDF in terms of its equilibrium configuration, I proceed with the calculation of the current. 

\section{Auxiliary expressions for the current density}
I introduce function $J_{\bk\eps}(\omega)=\textrm{Tr}[\hat{w}_{\bk\eps}]$. As it has been mentioned in the main text, there are two main contributions to this function which ultimately will contribute to the IFE.  The first one is given by the following expression:
\beg\label{Jife1}
\begin{split}
&J_{\bk\eps}^{(\textrm{I})}(\omega)=-\frac{8(e\alpha_{\textrm{so}})^2(\bn{\mathbf E})(\bn{\mathbf E}^*)}{m^2\omega^2\left[\left(\omega-\frac{\bk\bq}{m}\right)^2+\left(\frac{1}{\tau}\right)^2\right]}\sum\limits_{s=\pm}\frac{k^2(k-sm\alpha_{\textrm{so}})^2}{({z}_{q,-}^2+4\Delta_k^2)}\left(2F_{\bk\eps}^{(s)}-F_{\bk\eps+{\omega}}^{(s)}-F_{\bk\eps-{\omega}}^{(s)}\right).
\end{split}
\en
Expanding this expression up to the linear order in $\bq$, performing the remaining integrations and keeping only terms with required vector structure I find expression (\ref{j1}) listed in the main text.

There remains one more term with the desired vector structure which is described by the following expression:
\beg\label{Fifth}
\begin{split}
J_{\bk\eps}^{(\textrm{II})}(\omega)&=-\frac{8(e\alpha_{\textrm{so}})^2(\bn\times{\mathbf E})(\bn\times{\mathbf E}^*)}{\omega^2({z}_{q-}^2+4\Delta_k^2)({z}_{q+}^2+4\Delta_k^2)}\sum\limits_{s=\pm}\Delta_k^2\left(2F_{\bk\eps}^{(s)}-F_{\bk\eps+{\omega}}^{(s)}-F_{\bk\eps-{\omega}}^{(s)}\right).
\end{split}
\en 
In this expression we have neglected the contribution originating from $\hat{\gamma}_{\omega}^{(0)}$ since its contribution to the current will be at least of the order of $\omega/\veps_F$ smaller than the one from (\ref{Fifth}). Thus, performing the remaining integrations in (\ref{Fifth}) and expanding it up to the linear order in $v_Fq/\omega$ for the current density we find
\beg\label{j2}
\begin{split}
{\mathbf j}_{\textrm{IFE}}^{(\textrm{II})}(\bq,\omega)&=-\frac{5m}{\pi}\left(\frac{e\alpha_{\textrm{so}}}{\omega}\right)^3\left(\frac{\veps_F}{\omega}\right)\frac{\alpha_{\textrm{so}}}{\omega}B_{\omega\tau}\left\{(\bq{\mathbf E}_0){\mathbf E}_0^*+{\mathbf E}_0(\bq{\mathbf E}_0^*)\right\}.
\end{split}
\en
The dimensionless function $B_{\omega\tau}$ in (\ref{j2}) is defined according to
\beg\label{hwt}
B_{\omega\tau}=\frac{[1-4\zeta^2+(\omega\tau)^2](\omega\tau)^6}{[1+(\omega\tau-2\zeta)^2]^2[1+(\omega\tau+2\zeta)^2]^2}.
\en
It is easy to check that this contribution is by a factor of $m\alpha_{\textrm{so}}^2/\veps_F$ smaller than the first one (\ref{j1}).

\end{widetext}

\end{appendix}

\end{document}